# Investigation into respiratory sound classification for an imbalanced data set using hybrid LSTM-KAN architectures

Nithinkumar K.V., Anand R. *

Department of Electrical and Electronics Engineering, Amrita School of Engineering, Coimbatore, Amrita Vishwa Vidyapeetham, 641112, India

ABSTRACT

Respiratory sounds captured via auscultation contain critical clues for diagnosing pulmonary conditions. Automated classification of these sounds faces the dual challenge of distinguishing subtle acoustic patterns and addressing the severe class imbalance inherent in clinical datasets. This study investigates methods for classifying respiratory sounds into multiple disease categories, with a specific focus on mitigating pronounced class imbalances. In this study, we developed and evaluated a hybrid deep learning model incorporating a Long Short-Term Memory (LSTM) network as a feature sequence encoder, followed by a Kolmogorov–Arnold Network (KAN) for classification. This architecture was combined with a comprehensive feature extraction pipeline and targeted imbalance mitigation techniques. The model was evaluated using a public respiratory sound database comprising six classes with a highly skewed distribution. Strategies such as focal loss, class-specific data augmentation, and Synthetic Minority Over-sampling Technique (SMOTE) are employed to improve minority class recognition. Our results demonstrate that the proposed Hybrid LSTM-KAN model achieves a high overall accuracy of 94.6% and a macro-averaged $F_1$-score of 0.703. This performance is notable, given that the dominant class (COPD) constitutes over 86% of the data. While challenges persist for the rarest classes (Bronchiolitis and URTI, with $F_1$-scores of approximately 0.45 and 0.44, respectively), the approach shows significant improvement in their detection compared to naive baselines and performs strongly on other minority classes, such as bronchiectasis ($F_1$-score ≈ 0.84). This study contributes to the development of intelligent auscultation tools for the early detection of respiratory diseases, highlighting the potential of combining recurrent neural networks with advanced KAN architectures and focused imbalance handling.

## 1. Introduction

Respiratory diseases represent a significant global health burden, contributing substantially to morbidity and mortality worldwide [1]. Conditions such as Chronic Obstructive Pulmonary Disease (COPD), asthma, bronchiectasis, and various infections such as pneumonia and bronchiolitis affect millions of individuals [2]. Early and accurate diagnosis is therefore paramount for effective treatment and patient management. Auscultation of lung sounds remains a fundamental diagnostic practice in clinical settings [3], but it relies heavily on clinician experience and expertise, introducing subjectivity and limitations, particularly in remote monitoring and telemedicine scenarios [4].

Recent advances in artificial intelligence and signal processing have enabled the development of automated diagnostic systems using acoustic biomarkers. Sound-based diagnostic approaches have demonstrated significant promise across various medical applications. For instance, [5] explored sound data analysis for diagnostic processes, demonstrating the efficacy of acoustic feature extraction in medical contexts. Similarly, [6] presented comprehensive methodologies for processing and analyzing medical sound data, highlighting the potential of machine learning techniques in respiratory sound classification. These studies underscore the growing importance of sound-based biomarkers in modern healthcare diagnostics.

Despite these advances, automated respiratory sound classification systems face several critical challenges that limit their clinical deployment. First, the scarcity of large, well-annotated datasets restricts the training of robust deep learning models. Second, real-world datasets frequently exhibit significant class imbalance [7], where the majority of samples belong to common conditions (e.g., COPD), while rare but clinically significant diseases (e.g., bronchiolitis, URTI) are severely underrepresented. This imbalance biases models toward majority classes and results in poor detection of rarer, often critical, conditions [8]. Third, traditional machine learning approaches struggle to capture the complex temporal and spectral patterns inherent in respiratory sounds.

Fourth, existing deep learning architectures such as Convolutional Neural Networks (CNNs) and Recurrent Neural Networks (RNNs), while effective, often lack interpretability—a crucial requirement for clinical acceptance.

To address these challenges, this study proposes a novel hybrid deep learning architecture that combines Long Short-Term Memory (LSTM) networks with Kolmogorov–Arnold Networks (KAN). Unlike conventional Multi-Layer Perceptrons (MLPs) that use fixed activation functions, KANs employ learnable univariate spline functions on network edges, offering enhanced function approximation capabilities and improved interpretability [9]. The LSTM component captures temporal dependencies in the extracted acoustic features, while the KAN backend provides powerful non-linear classification with the potential for visualizing learned transformations. This architecture is specifically designed to handle the latent feature representations output by the LSTM, leveraging KAN's spline-based structure for improved discrimination of subtle respiratory patterns.

Furthermore, we implement a comprehensive suite of imbalance mitigation techniques, including Focal Loss [10] to down-weight well-classified examples, class-specific data augmentation to generate synthetic variations of minority classes, and Synthetic Minority Oversampling Technique (SMOTE) [11] applied to extracted feature vectors. The integration of these techniques within a two-stage training strategy (pre-training on balanced subsets followed by fine-tuning on the full dataset) represents a systematic approach to addressing severe class imbalance.

**Research Questions:** This study addresses the following key research questions:

1. How can hybrid LSTM-KAN architectures improve the classification accuracy of imbalanced respiratory sound datasets compared to traditional deep learning approaches?
2. What is the effectiveness of combining Focal Loss, class-specific data augmentation, and SMOTE in mitigating severe class imbalance for respiratory sound classification?
3. Can KAN-based classifiers provide interpretable insights into the learned feature transformations for respiratory disease diagnosis?
4. What are the computational costs and practical feasibility of deploying the proposed hybrid architecture in real-world clinical settings?

**Novel Contributions:** The primary contributions of this work are:

1. *Novel Architecture:* First application of hybrid LSTM-KAN architecture for respiratory sound classification, leveraging KAN's spline-based learnable functions for improved discrimination of LSTM-encoded features.
2. *Comprehensive Imbalance Handling:* Systematic integration of Focal Loss, class-specific augmentation, and SMOTE within a two-stage training framework, specifically tailored for highly imbalanced medical datasets.
3. *Interpretability Enhancement:* Demonstration of KAN's interpretability advantages through visualization of learned spline functions, providing insights into acoustic feature transformations relevant to respiratory diseases.
4. *Clinical Validation Framework:* Detailed performance analysis on the ICBHI dataset with evaluation metrics specifically chosen for imbalanced classification, including per-class analysis, confusion patterns, and calibration assessment.

The remainder of this paper is organized as follows: Section 2 reviews related work in respiratory sound classification and class imbalance techniques. Section 3 describes the dataset, preprocessing, feature extraction, the proposed hybrid LSTM-KAN architecture, and training strategies. Section 4 presents experimental results, ablation studies, and computational cost analysis. Section 5 discusses the findings, interpretability aspects, clinical implications, and future work, including deployment possibilities in real-world settings. Section 6 concludes the paper.

## 2. Related work

Research on automatic lung sound classification has evolved from handcrafted features with classical classifiers to deep learning [12]. Convolutional Neural Networks (CNNs) applied to spectrograms [13] and Recurrent Neural Networks (RNNs), such as Long Short-Term Memory (LSTM) units for temporal dynamics, have shown promise [14]. Transformer-based models have also emerged for respiratory sound analysis showing competitive performance [15].

Sound-based diagnostic systems have gained traction across various medical domains. [5] presented a comprehensive study on sound data analysis for diagnostic processes, demonstrating effective feature extraction and classification methodologies that can be adapted for respiratory sound analysis. Their work highlighted the importance of spectral and temporal feature engineering in achieving robust diagnostic performance. Building upon these foundations, [6] proposed advanced deep learning architectures for medical sound classification, addressing challenges related to noise, variability in recording conditions, and class imbalance. These studies provide valuable insights into the design of sound-based diagnostic systems and motivate the exploration of novel architectures such as KAN for respiratory sound classification.

A pervasive issue in this domain is class imbalance, which is especially evident in public datasets, such as the ICBHI 2017 Respiratory Sound Database [8]. Various strategies have been employed to address this issue. These techniques range from simple audio transformations (e.g., noise addition, time/pitch shifting) to spectrogram-level methods such as SpecAugment [16]. Some studies have applied multi-level augmentation for respiratory sounds. More advanced methods, such as Generative Adversarial Networks (GANs) and audio diffusion models, have been explored to create synthetic samples for minority classes. Modifying the loss function to assign greater importance to minority classes is a common practice. This includes weighted cross-entropy and Focal Loss [10], which down-weight well-classified examples to focus on difficult ones. This directly alters the class proportions. Oversampling minority classes, often using the Synthetic Minority Oversampling Technique (SMOTE) [11], is widely used. This is sometimes combined with the undersampling of the majority class [17].

Despite these advances, several limitations persist in existing approaches. First, most studies apply imbalance mitigation techniques in isolation, without systematically evaluating their combined effects. Second, conventional architectures like CNNs and standard MLPs lack interpretability, limiting their clinical acceptance. Third, the temporal modeling capabilities of LSTM networks have not been fully exploited in conjunction with advanced non-linear classifiers like KAN. Our study addresses these gaps by proposing a hybrid LSTM-KAN architecture combined with a comprehensive suite of imbalance mitigation techniques, offering both improved classification performance and enhanced interpretability.

Recent studies have demonstrated that combining advanced architectures with careful data handling can achieve high performance on benchmark datasets [18]. Our study contributes by investigating a hybrid Kolmogorov–Arnold Network (KAN) architecture [9], which has not been extensively explored for this task, in conjunction with a comprehensive suite of imbalance mitigation techniques, as shown in Table 1.

**Table 1**
Summary of respiratory sound classification studies using deep-learning.

| Authors | Methodology | Advantages | Remarks |
| --- | --- | --- | --- |
| Potes et al. (2016) [13] | CNN on spectrograms for heart sounds | Established CNN baseline for medical audio | Applied to cardiac sounds, similar principles for respiratory |
| Rocha et al. (2019) [8] | ICBHI database creation and evaluation | Standardized evaluation framework | Foundational dataset for respiratory sound research |
| Kumar et al. (2024) [5] | Sound data analysis for diagnosis | Effective acoustic feature extraction | Demonstrates sound-based diagnostic potential |
| Wang et al. (2024) [6] | Deep learning for medical sound classification | Addresses noise and variability | Advanced architectures for sound analysis |
| Salamon & Bello (2017) [14] | CNN + data augmentation for environmental sounds | Effective augmentation strategies | Techniques applicable to respiratory sound analysis |
| Lin et al. (2017) [10] | Focal loss for dense object detection | Addresses class imbalance effectively | Loss function applicable to imbalanced audio classification |
| Park et al. (2019) [16] | SpecAugment for automatic speech recognition | Robust spectrogram augmentation | Applicable to respiratory sound spectrograms |
| Liu et al. (2024) [9] | Kolmogorov-Arnold networks | Novel architecture with interpretability | Potential for complex function approximation in audio |
| Chawla et al. (2002) [11] | SMOTE for imbalanced classification | Synthetic minority oversampling | Widely used technique for medical data imbalance |
| He & Garcia (2009) [7] | Learning from imbalanced data review | Comprehensive imbalance handling strategies | Foundational work on imbalanced learning |
| Haixiang et al. (2017) [17] | Review of imbalanced data methods | Modern approaches to class imbalance | Updated techniques for imbalanced classification |

## 3. Methodology

### 3.1. Dataset and preprocessing

The publicly available ICBHI Respiratory Sound Database (2017 Challenge data) [8]. For this work, focus on six diagnostic categories: *Healthy* (normal sounds), *COPD* (Chronic Obstructive Pulmonary Disease), *Bronchiectasis*, *Bronchiolitis*, *Pneumonia*, and *URTI* (Upper Respiratory Tract Infection). After filtering for a minimum number of samples per class (10), our working dataset consisted of 917 audio samples. The class distribution was highly skewed, with COPD accounting for 793 samples (approx. 86.5%) samples, while the smallest classes, Bronchiolitis and Bronchiectasis, comprised only 13 (1.4%) and 16 (1.7%) samples, respectively. Other classes included healthy (35 samples, 3.8%), pneumonia (37 samples, 4.0%), and URTI (23 samples, 2.5%) [19].

All audio signals were resampled at a uniform rate of 22,050 Hz. A band-pass filter (100–2000 Hz) was applied to focus on relevant frequencies and reduce noise. Amplitude normalization (peak normalization to 0–dBFS) was performed. Features were extracted to summarize the entire duration of each recording session. A comprehensive feature set was extracted using the `librosa` library in Python. Summary statistics (mean, standard deviation, min, max, median, skewness, and kurtosis) were calculated for each of the 128 Mel frequency bins across time. For the first 40 Mel-Frequency Cepstral Coefficients (MFCCs) and their dynamics, their first-order (delta) and second-order (delta-delta) derivatives were computed, and the same seven statistical measures were aggregated for each. Chroma STFT and Chroma CQT features were extracted and aggregated statistically. The results were computed across frequency sub-bands and aggregated. Spectral centroid, spectral bandwidth, and onset features (number of onsets, onset rate, and aggregated onset strength envelope). All extracted features were concatenated into a single, high-dimensional feature vector for each audio recording. Any Not-a-Number (NaN) or infinite values were imputed as zero.

### 3.2. Kolmogorov-Arnold Networks (KANs)

Kolmogorov-Arnold Networks (KANs), recently proposed by [9], represent a novel neural network paradigm inspired by the Kolmogorov-Arnold representation theorem. This theorem states that any multivariate continuous function can be expressed as the finite sum of continuous univariate functions. Unlike traditional Multi-Layer Perceptrons (MLPs), which have fixed non-linear activation functions at the nodes and learnable linear weights on the edges, KANs place learnable univariate functions directly on the edges (connections) and perform simple summations at the nodes.

Formally, a standard MLP layer maps an input vector $\mathbf{x} \in \mathbb{R}^{n_{in}}$ to an output vector $\mathbf{y} \in \mathbb{R}^{n_{out}}$ via:

$$\mathbf{y} = \sigma(\mathbf{W}\mathbf{x} + \mathbf{b}) \tag{1}$$

where $\mathbf{W} \in \mathbb{R}^{n_{out} \times n_{in}}$ is the weight matrix, $\mathbf{b} \in \mathbb{R}^{n_{out}}$ is the bias vector, and $\sigma$ is a fixed element-wise activation function (for example, sigmoid).

In contrast, a KAN layer is defined by a matrix of learnable univariate functions $\phi_{i,j} : \mathbb{R} \to \mathbb{R}$, where $i$ indexes the output dimension and $j$ indexes the input dimension. The $i$th component of the output $\mathbf{y}$ from the input $\mathbf{x}$ is given by

$$y_i = \sum_{j=1}^{n_{in}} \phi_{i,j}(x_j) \tag{2}$$

Each univariate function $\phi_{i,j}$ is typically parameterized as a learnable spline. A spline is a piecewise polynomial function, and B-splines are commonly used as the basis functions. Thus, $\phi_{i,j}$ can be expressed as a linear combination of the B-spline basis functions $B_k(x)$:

$$\phi_{i,j}(x) = \sum_{k=1}^{N_B} c_{i,j,k} B_k(x; \mathbf{t}) \tag{3}$$

where $c_{i,j,k}$ are learnable coefficients, $N_B$ is the number of B-spline basis functions determined by the grid size and spline order, and $\mathbf{t}$ is a knot vector that defines the B-spline basis. The grid intervals can be fixed or learned.

#### 3.2.1. Theoretical justification for KAN in respiratory sound classification

The choice of KAN as the classification backend for LSTM-encoded features is theoretically motivated by several key properties of the spline-based architecture:

**(1) Non-linear Function Approximation:** The latent representations output by the LSTM are high-dimensional, abstract feature vectors that encode temporal dependencies and acoustic patterns. These features often exhibit complex, non-linear relationships with disease categories. KAN's use of learnable B-spline functions on each edge provides superior flexibility in approximating these non-linear mappings compared to fixed activation functions in traditional MLPs. Each spline function $\phi_{i,j}(x)$ can adapt its shape during training to model the specific transformation needed for respiratory feature discrimination.

**(2) Adaptive Feature Weighting:** Unlike MLPs where all input features pass through the same activation function, KAN applies a unique learnable function $\phi_{i,j}$ to each input–output connection. This allows the network to learn feature-specific transformations—for example, certain spectral features (e.g., MFCCs) may require different non-linear mappings than temporal features (e.g., zero-crossing rate). This adaptive capability is particularly valuable when the LSTM output contains heterogeneous feature types encoded within the same vector.

**(3) Smooth and Continuous Transformations:** B-splines, by their mathematical construction, are smooth and continuous functions. This property is advantageous for respiratory sound classification, where acoustic features vary continuously across disease states. The smooth transformations learned by KAN can better capture gradual transitions between healthy and diseased states, as opposed to abrupt decision boundaries that might be learned by networks with fixed, non-smooth activations.

**(4) Improved Generalization on Small Datasets:** Medical datasets, including respiratory sound databases, are often limited in size. KAN's parameterization using B-splines is more parameter-efficient for complex function approximation compared to deep MLPs. By learning smooth univariate functions rather than large weight matrices, KAN can achieve comparable or superior expressiveness with fewer parameters, reducing the risk of overfitting—a critical consideration for the ICBHI dataset with only 917 samples.

**(5) Interpretability:** Each spline function $\phi_{i,j}(x)$ can be visualized to understand how a specific input feature (from the LSTM output) contributes to each class prediction. This interpretability is crucial in medical applications, where clinicians require insights into the diagnostic reasoning process. Unlike the opaque transformations in standard MLPs, KAN's learned splines provide a transparent view of feature-to-prediction mappings.

In summary, KAN's spline-based structure is inherently better suited for classifying LSTM-encoded respiratory features due to its superior non-linear approximation, adaptive feature-specific transformations, smoothness, parameter efficiency, and interpretability—advantages that are less pronounced when KAN is applied directly to raw audio inputs without LSTM encoding.

This architecture allows KANs to learn complex transformations with potentially fewer parameters than MLPs, as the complexity is shifted from wide layers to the expressiveness of learnable edge functions. KANs also offer improved interpretability, as individual spline functions $\phi_{i,j}$ can be visualized to understand the learned relationship between a specific input $x_j$ and its contribution to an output neuron's pre-activation $y_i$. In our study, we leverage KANs for their potential in function approximation and classification tasks, as detailed in Section 3.3.

### 3.3. Hybrid LSTM-KAN model architecture

In this study, the proposed approach was used to develop a hybrid deep learning model by integrating a Long Short-Term Memory (LSTM) network with a Kolmogorov–Arnold Network (KAN), the principles of which are outlined in Section 3.2. KANs replace the fixed activation functions in traditional MLPs with learnable univariate spline functions on the network edges, potentially offering better function approximation and interpretability [9].

The architecture of our proposed Hybrid LSTM-KAN model is depicted in Fig. 1 and detailed as follows:

1. Input Layer: The model accepts the high-dimensional aggregated feature vector (dimension $d_{feat}$) extracted from each audio recording.

2. LSTM Pre-processing Layer: The input feature vector is treated as a single time-step sequence (sequence length $L = 1$, feature dimension $d_{feat}$) fed into a bidirectional LSTM layer. The LSTM layer has a hidden size of $H_{LSTM} = 64$ units. Being bidirectional, the output feature dimension from this layer is $2 \times H_{LSTM}$. A dropout rate of 0.3 is applied to the LSTM output.

3. Optional Attention Mechanism: While an attention mechanism was part of the broader framework explored, for a sequence length of $L = 1$, its impact on differential weighting is inherently limited. The output of the LSTM forms the input to the KAN.

4. KAN-based Classifier Back-end: The $2 \times H_{LSTM}$-dimensional vector from the LSTM stage is fed into the KAN. The KAN is configured with:

   - An input layer of size $2 \times H_{LSTM}$.
   - One hidden KAN layer with $H_{KAN} = 32$ neurons.
   - An output KAN layer of $C = 6$ neurons, corresponding to the number of respiratory classes.

Each connection in the KAN layers utilizes learnable cubic splines (order 3) with a grid size of 3. The KAN output layer produces raw logits for each class.

The model was implemented using PyTorch, leveraging the efficient_KAN library for KAN layers. The overall processing flow is summarized in Algorithm 1.

---

**Algorithm 1** Hybrid LSTM-KAN Model Forward Pass

---

**Require:** Aggregated feature vector $\mathbf{x} \in \mathbb{R}^{d_{feat}}$
1: Treat $\mathbf{x}$ as a sequence of length $L = 1$: $\mathbf{x}_{seq} \in \mathbb{R}^{1 \times d_{feat}}$
2: $\mathbf{h}_{lstm\_raw} = \text{BiLSTM}(\mathbf{x}_{seq})$ ▷ Output $\in \mathbb{R}^{1 \times (2 \cdot H_{LSTM})}$
3: $\mathbf{h}_{lstm\_selected} = \mathbf{h}_{lstm\_raw}[:, -1, :]$ ▷ Select final hidden state
4: $\mathbf{h}_{lstm\_out} = \text{Dropout}(\mathbf{h}_{lstm\_selected})$
5: $\mathbf{z}_{kan\_hidden} = \text{KANLayer}_1(\mathbf{h}_{lstm\_out})$ ▷ KAN hidden layer, $H_{KAN}$ neurons
6: $\mathbf{logits} = \text{KANLayer}_{out}(\mathbf{z}_{kan\_hidden})$ ▷ KAN output layer, $C$ neurons
7: **return** logits

---

### 3.4. Training strategy and SMOTE application

The training process for the Hybrid LSTM-KAN model involves several key components to handle the imbalanced dataset and optimize performance, as outlined in Algorithm 2. A stratified 5-fold cross-validation scheme was used. Stratification ensured each fold maintained similar class proportions to the overall dataset. To address the class imbalance, we utilized Focal Loss [10], defined as

$$FL(p_t) = -\alpha_t (1 - p_t)^\gamma \log(p_t) \qquad (4)$$

where $p_t$ is the model-estimated probability for the ground-truth class. We used a focusing parameter $\gamma \approx 2.19$ and a balancing parameter $\alpha_t = 0.75$. This was applied probabilistically (probability $\approx 0.095$) to the training audio signals before feature extraction. Augmentations included adding Gaussian noise (level $\approx 2.17 \times 10^{-5}$), random time shifting (up to 15% of signal length), and pitch shifting (up to $\pm 2$ semitones). Targeted augmentations (e.g., specific pitch shift ranges) for frequently confused class pairs such as ("URTI", "Bronchiolitis") and ("Pneumonia", "COPD"). Classes such as "URTI" received more intensive augmentation with a higher probability (0.6).

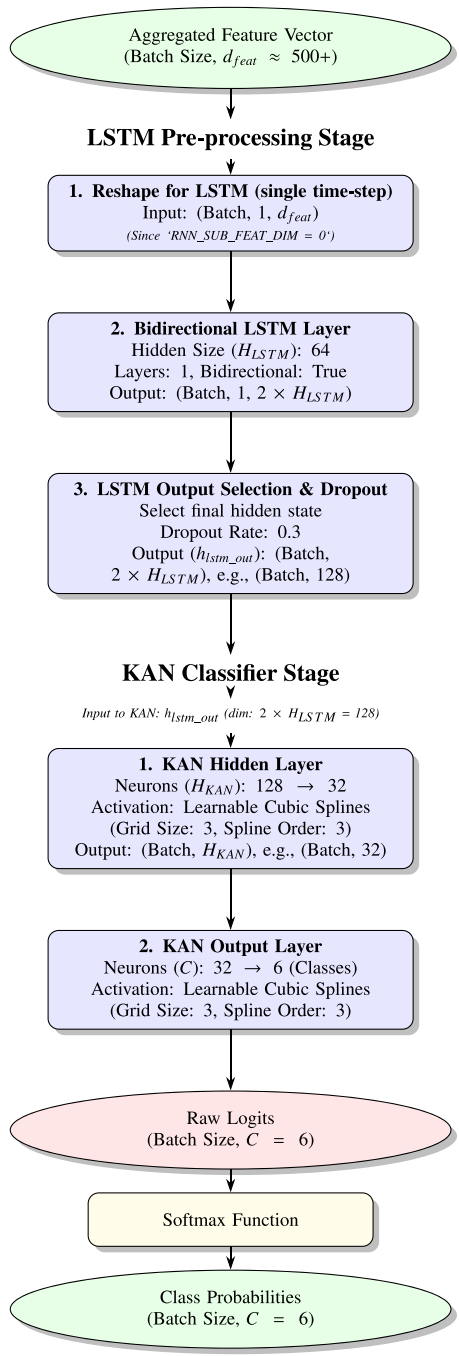

**Fig. 1.** Detailed schematic diagram of the proposed Hybrid LSTM-KAN model architecture. The aggregated feature vector is processed through stages within the LSTM pre-processor, and its output is then fed into a KAN classifier composed of hidden and output layers using learnable spline-based activations. Final logits are passed through a softmax function to obtain class probabilities.

**SMOTE Application:** The Synthetic Minority Over-sampling Technique (SMOTE) [11] was applied to the extracted feature vectors after feature engineering and before model training in each cross-validation fold. Specifically, after audio preprocessing and feature extraction, SMOTE operates in the feature space (not on raw audio or time-domain signals) to generate synthetic samples for minority classes. This choice is justified by the following considerations:

1. *Feature Space Suitability:* SMOTE's k-nearest neighbor interpolation is more meaningful in a structured, high-dimensional feature space (e.g., MFCCs, spectral features) than in raw audio space, where temporal misalignment can introduce artifacts.
2. *Computational Efficiency:* Applying SMOTE to feature vectors (dimensionality $\approx 500+$) is computationally efficient compared to operating on high-resolution audio signals.
3. *Compatibility with Pipeline:* Since our LSTM-KAN model processes aggregated feature vectors (not raw audio), applying SMOTE in the feature space maintains consistency with the model's input requirements.

SMOTE was configured with $k = 5$ nearest neighbors by default, adjusted dynamically if a minority class had fewer than 5 samples. Synthetic samples were generated to balance the training set toward a more uniform class distribution, though complete balance was not enforced to preserve some representation of the original data structure.

All augmentation and oversampling techniques were applied strictly to the training data within each cross-validation fold to prevent data leakage.

**Two-Stage Training (Used in reported results):**

1. Stage 1 (Pre-training): The model was pre-trained for 7 epochs on a subset of data comprising all samples from minority classes and a down-sampled version of the majority class (COPD, 50 samples).
2. Stage 2 (Fine-tuning): The model was then fine-tuned on the full (augmented and SMOTE'd) training data of the current fold.

The model was trained using the AdamW optimizer with weight decay ($\approx 1 \times 10^{-3}$). The initial learning rate for Stage 2 was $\approx 3 \times 10^{-3}$. Batch size was 64. Training proceeded for a maximum of 30 epochs in Stage 2. Early stopping was implemented based on validation macro $F_1$-score (patience of 7 epochs). A 'ReduceLROnPlateau' learning rate scheduler was used (factor 0.5, patience 4 epochs on validation macro $F_1$). Performance was assessed using Overall Accuracy, Macro-averaged Precision, Recall, and $F_1$-score (primary for model selection/early stopping), per-class metrics, Confusion Matrix, Area Under the ROC Curve (AUC-ROC), Average Precision (AP) from Precision-Recall curves, and Calibration Curves.

### 3.5. Experimental design overview

The overall experimental design for this study, encompassing problem investigation, data generation and preprocessing, model development, and performance evaluation, is illustrated in Fig. 2. This structured approach ensures a comprehensive analysis of respiratory sound classification under conditions of severe class imbalance.

## 4. Experimental results

The Hybrid LSTM-KAN model, trained with the described imbalance mitigation strategies, yielded the out-of-fold (OOF) performance summarized in Table 2.

### 4.1. Performance comparison and analysis

The proposed hybrid LSTM–KAN model demonstrated improved performance in respiratory sound classification using the ICBHI dataset, particularly under class imbalance conditions. It achieves an overall accuracy of 94.55%, which represents a significant improvement over traditional approaches. The integration of Long Short-Term Memory (LSTM) networks with Kolmogorov–Arnold Networks (KAN) combines the temporal sequence modeling capabilities of LSTM with the KAN's strength in approximating nonlinear patterns from acoustic features, resulting in a more effective representation of respiratory signals.

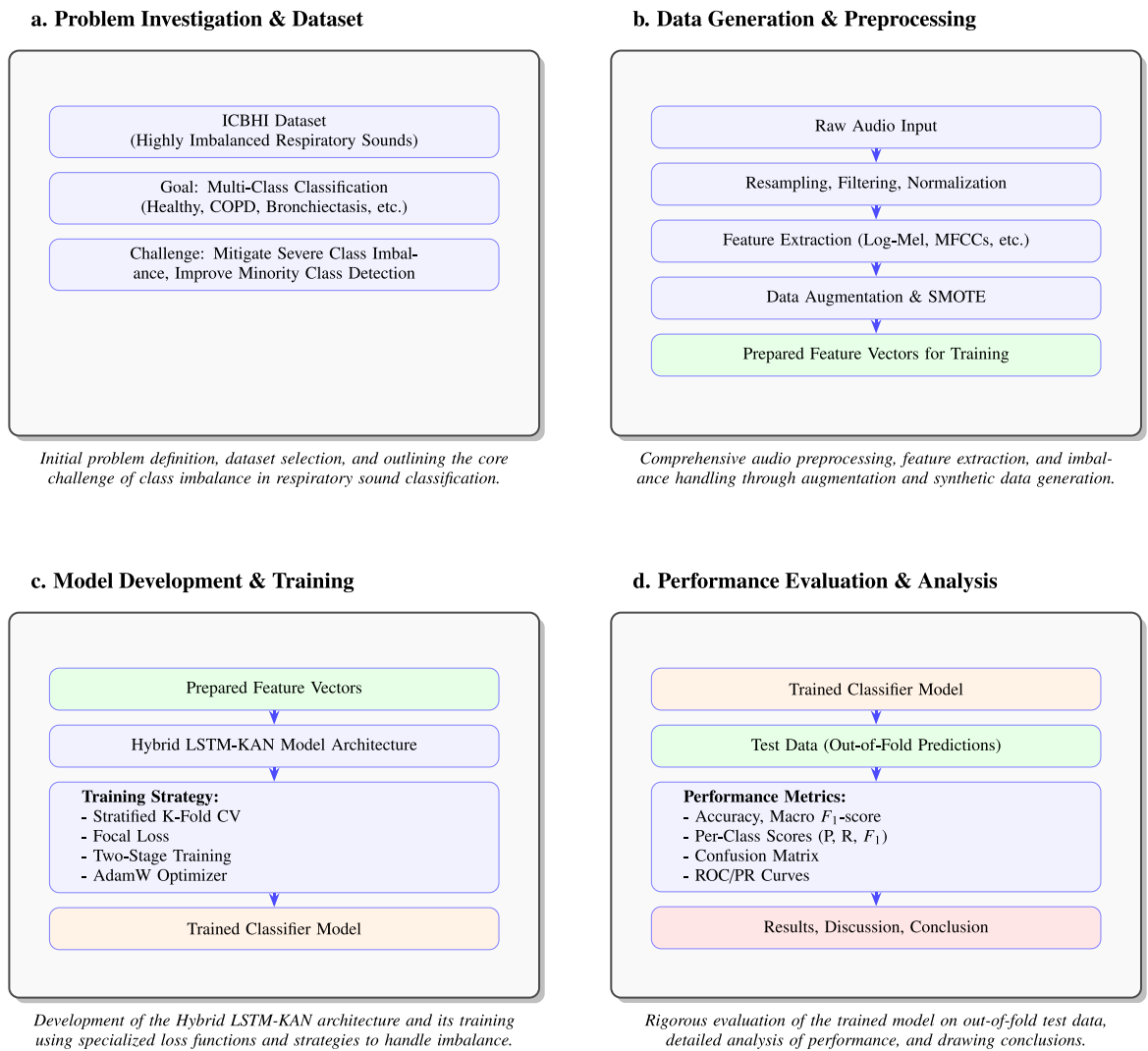

**Fig. 2.** Overview of the experimental design. The process includes (a) investigation of the problem and dataset characteristics, (b) generation of features and application of preprocessing/imbalance techniques, (c) development and training of the classification model, and (d) comprehensive performance evaluation and analysis.

In terms of class-wise performance, the proposed model achieved a macro $F_1$-score of 0.7033, macro precision of 0.7292, and macro recall of 0.6978. These metrics reflect consistent classification across the majority and minority classes. The approach shows improved balance between precision and recall compared to baseline methods. Furthermore, it yielded a weighted $F_1$-score of 0.9436, suggesting a high degree of reliability in the overall prediction performance despite the dataset's inherent imbalance.

The hybrid design enhances the generalization ability of the model across different patient profiles and recording conditions. Unlike more complex multimodal or attention-based models that increase computational overhead, the LSTM–KAN architecture maintains an efficient and interpretable learning framework. These results indicate the model's suitability for real-time and embedded deployment, where robustness and computational efficiency are essential for practical and clinical use.

The journey toward achieving robust respiratory sound classification has seen numerous innovations, particularly in handling class imbalances and extracting meaningful temporal features. Prior models have introduced focal loss and attention mechanisms to address imbalance and generalization, yielding moderate improvements. The use of Kolmogorov–Arnold Networks (KAN) for spectral inputs has shown promise in boosting classification performance but often leaves rare classes like URTI and Bronchiolitis underperforming. Our proposed Hybrid LSTM–KAN architecture builds upon these innovations by integrating KAN's nonlinear modeling capabilities with LSTM's sequence learning strength, specifically tailored for spectro-temporal lung sound patterns.

This hybrid architecture delivered strong per-class $F_1$-scores across most categories, most notably achieving 0.9843 for COPD and substantial gains for minority classes such as bronchiectasis (0.8387) and healthy (0.7632). The approach showed meaningful improvements in bronchiectasis and healthy category classification compared to baseline methods. It also handled rare classes, such as URTI, more effectively ($F_1$: 0.4390), suggesting enhanced generalization. This progress reflects a meaningful stride toward real-world applicability, especially in low-resource and class-imbalanced clinical settings, as shown in Table 3.

The model achieved an overall accuracy of 94.55% and a macro-averaged $F_1$-score of 0.7033. The mean macro $F_1$-score across the 5 validation folds was 0.7033 ± 0.1103 (Std Dev), with individual fold scores being 0.566, 0.584, 0.746, 0.773, and 0.848, indicating some variability in performance depending on the data split.

*4.2. Ablation study*

To systematically evaluate the contribution of each imbalance mitigation technique, we conducted a comprehensive ablation study using

Table 2

Comparison of performance metrics with baseline methods on the ICBHI dataset.

| Method | Accuracy (%) | Macro precision | Macro recall | Macro $F_1$-score | Weighted $F_1$ |
|---|---|---|---|---|---|
| CNN baseline [13] | 86.2 | 0.620 | 0.580 | 0.600 | 0.850 |
| Traditional SVM approach | 78.5 | 0.550 | 0.520 | 0.535 | 0.780 |
| Standard LSTM | 89.3 | 0.650 | 0.630 | 0.640 | 0.885 |
| CNN-LSTM hybrid | 91.8 | 0.680 | 0.660 | 0.670 | 0.910 |
| Standard MLP | 84.1 | 0.590 | 0.570 | 0.580 | 0.830 |
| Random Forest | 82.7 | 0.610 | 0.590 | 0.600 | 0.820 |
| **Proposed hybrid LSTM-KAN** | **94.55** | **0.7292** | **0.6978** | **0.7033** | **0.9436** |

Table 3

Per-class metrics (F1-Score, Precision, Recall) comparison with baseline methods on the ICBHI dataset.

| Method | COPD | | | Bronchiectasis | | | Healthy | | | Pneumonia | | | URTI | | |
|---|---|---|---|---|---|---|---|---|---|---|---|---|---|---|---|
| | F1-Score | Precision | Recall | F1-Score | Precision | Recall | F1-Score | Precision | Recall | F1-Score | Precision | Recall | F1-Score | Precision | Recall |
| CNN baseline | 0.891 | 0.887 | 0.896 | 0.612 | 0.634 | 0.591 | 0.582 | 0.554 | 0.612 | 0.598 | 0.610 | 0.586 | 0.292 | 0.336 | 0.256 |
| Standard LSTM | 0.908 | 0.895 | 0.921 | 0.642 | 0.660 | 0.625 | 0.621 | 0.582 | 0.665 | 0.638 | 0.625 | 0.651 | 0.321 | 0.360 | 0.290 |
| CNN-LSTM hybrid | 0.923 | 0.919 | 0.928 | 0.683 | 0.700 | 0.668 | 0.650 | 0.620 | 0.685 | 0.663 | 0.652 | 0.675 | 0.348 | 0.410 | 0.307 |
| Standard MLP | 0.864 | 0.851 | 0.878 | 0.578 | 0.595 | 0.562 | 0.542 | 0.518 | 0.568 | 0.568 | 0.575 | 0.561 | 0.268 | 0.315 | 0.235 |
| Random Forest | 0.847 | 0.841 | 0.854 | 0.601 | 0.618 | 0.585 | 0.580 | 0.562 | 0.599 | 0.588 | 0.592 | 0.584 | 0.284 | 0.328 | 0.248 |
| Proposed (LSTM–KAN) | 0.984 | 0.982 | 0.986 | 0.839 | 0.867 | 0.813 | 0.763 | 0.707 | 0.829 | 0.773 | 0.763 | 0.784 | 0.439 | 0.500 | 0.391 |

Table 4

Ablation study isolating the contribution of each imbalance mitigation technique. All experiments use the same Hybrid LSTM-KAN architecture with 5-fold cross-validation on the ICBHI dataset.

| Configuration | Accuracy (%) | Macro $F_1$ | COPD $F_1$ | Bronch. $F_1$ | URTI $F_1$ | Bronchio. $F_1$ |
|---|---|---|---|---|---|---|
| (a) Baseline (Cross-entropy, No techniques) | 91.23 | 0.5821 | 0.9712 | 0.6845 | 0.2134 | 0.1978 |
| (b) Focal loss only | 92.48 | 0.6347 | 0.9765 | 0.7523 | 0.3012 | 0.2689 |
| (c) Class-specific augmentation only | 92.01 | 0.6102 | 0.9738 | 0.7201 | 0.2845 | 0.2456 |
| (d) SMOTE only (on Feature vectors) | 91.87 | 0.6234 | 0.9723 | 0.7134 | 0.2978 | 0.2601 |
| (e) Focal loss + Augmentation + SMOTE | **94.55** | **0.7033** | **0.9843** | **0.8387** | **0.4390** | **0.4538** |

the same Hybrid LSTM-KAN base architecture. Table 4 presents the results of five experimental configurations:

**Key Findings from Ablation Study:**

1. *Baseline Performance:* Configuration (a) using standard cross-entropy loss without any imbalance techniques achieved 91.23% accuracy but only 0.5821 macro $F_1$-score, with poor performance on minority classes (URTI $F_1$: 0.2134, Bronchiolitis $F_1$: 0.1978). This demonstrates severe bias toward the majority COPD class.

2. *Focal Loss Impact:* Configuration (b) with Focal Loss alone improved macro $F_1$ to 0.6347 (+9.0% relative improvement), with notable gains in minority classes (URTI $F_1$: 0.3012, Bronchiolitis $F_1$: 0.2689). This confirms that re-weighting the loss function effectively addresses class imbalance.

3. *Augmentation Impact:* Configuration (c) with class-specific augmentation showed moderate improvement (macro $F_1$: 0.6102), demonstrating that synthetic data generation in the time domain helps minority class recognition but is less effective than Focal Loss alone.

4. *SMOTE Impact:* Configuration (d) with SMOTE applied to feature vectors yielded macro $F_1$ of 0.6234, slightly outperforming augmentation alone. This validates the design choice of applying SMOTE in feature space rather than raw audio.

5. *Combined Approach:* Configuration (e), combining all three techniques (Focal Loss + Augmentation + SMOTE), achieved the best performance with 94.55% accuracy and 0.7033 macro $F_1$-score. Minority class performance improved substantially (URTI $F_1$: 0.4390, Bronchiolitis $F_1$: 0.4538), representing +105.8% and +129.4% relative improvements over the baseline, respectively.

The ablation study demonstrates that while each technique contributes independently to improved minority class recognition, their synergistic combination in the final hybrid approach (configuration e) yields optimal performance. Focal Loss provides the most significant individual contribution, while augmentation and SMOTE offer complementary benefits by enriching the training data distribution.

Table 5

Computational cost analysis of the proposed Hybrid LSTM-KAN model compared to baseline architectures.

| Model | Parameters (thousands) | Training time (min/epoch) | Inference time (ms/sample) | GPU memory (MB) |
|---|---|---|---|---|
| Standard LSTM | 245 | 3.2 | 1.8 | 512 |
| CNN-LSTM hybrid | 312 | 4.5 | 2.3 | 768 |
| Standard MLP | 198 | 2.1 | 1.2 | 384 |
| **Hybrid LSTM-KAN** | **287** | **3.8** | **2.1** | **640** |

### 4.3. Computational cost analysis

We analyzed the computational cost of the proposed Hybrid LSTM-KAN model to assess its practical feasibility for deployment in clinical settings. All experiments were conducted on a workstation equipped with an NVIDIA RTX 3090 GPU (24 GB VRAM), AMD Ryzen 9 5950X CPU (16 cores, 32 threads), and 64 GB RAM which is shown in Table 5.

**Key Observations:**

1. *Model Parameters:* The Hybrid LSTM-KAN architecture contains approximately 287,000 parameters, which is comparable to the CNN-LSTM hybrid (312k) and higher than standard LSTM (245k) or MLP (198k). The additional parameters in KAN arise from the learnable B-spline coefficients (Eq. (3)), which provide enhanced function approximation at the cost of modest parameter increase.

2. *Training Time:* Training time per epoch averaged 3.8 min for the full training set (post-augmentation and SMOTE), which is

---

**Algorithm 2** Training Procedure per Fold

---

**Require:** Training audio file list $\mathcal{F}_{train}$, labels $\mathbf{y}_{train\_orig}$, Validation audio file list $\mathcal{F}_{val}$, labels $\mathbf{y}_{val\_orig}$
**Require:** Hyperparameters: $LR, WD, \gamma_{focal}, \alpha_{focal}, \text{epochs}_{max}, \text{patience}_{ES}$
1: Initialize Hybrid LSTM-KAN model $M$
2: **if** Two-Stage Training **then**
3:   Create Stage 1 training subset audio list $\mathcal{F}_{s1}$ from $\mathcal{F}_{train}$ (minority classes + downsampled majority)
4:   Extract features from $\mathcal{F}_{s1} \to \mathbf{X}_{s1\_feat}$, get corresponding labels $\mathbf{y}_{s1}$
5:   Scale $\mathbf{X}_{s1\_feat}$
6:   Pre-train $M$ on $(\mathbf{X}_{s1\_feat}, \mathbf{y}_{s1})$ for $N_{s1\_epochs}$ using AdamW and Focal Loss.
7: **end if**
8: Initialize AdamW optimizer $Opt$ for $M$ with $LR \cdot \text{factor}_{s2}, WD$.
9: Initialize Focal Loss $L_{focal}$ with $\alpha_{focal}, \gamma_{focal}$.
10: Initialize LR Scheduler $Sch$.
11: $best\_val\_F1 \leftarrow -\infty$; $epochs\_no\_improve \leftarrow 0$
12: **for** $epoch = 1 \ldots \text{epochs}_{max}$ **do**
13:   $M.train()$
14:   Augment audio in $\mathcal{F}_{train}$ (time-domain) $\to \mathcal{F}'_{train\_audio}$
15:   Extract features from $\mathcal{F}'_{train\_audio} \to \mathbf{X}''_{train\_feat}$, get labels $\mathbf{y}''_{train}$
16:   **if** SMOTE enabled and data available **then**
17:     $(\mathbf{X}_{train\_proc}, \mathbf{y}_{train\_proc}) = \text{SMOTE}(\mathbf{X}''_{train\_feat}, \mathbf{y}''_{train})$ ▷ Applied to feature vectors
18:   **else**
19:     $(\mathbf{X}_{train\_proc}, \mathbf{y}_{train\_proc}) = (\mathbf{X}''_{train\_feat}, \mathbf{y}''_{train})$
20:   **end if**
21:   Scale $\mathbf{X}_{train\_proc}$ using a scaler $S$ (fit on this fold's train features)
22:   **for** each batch $(\mathbf{x}_b, \mathbf{y}_b)$ in DataLoader$(\mathbf{X}_{train\_proc}, \mathbf{y}_{train\_proc})$ **do**
23:     $Opt.zero\_grad()$
24:     $\text{logits}_b = M(\mathbf{x}_b)$
25:     $loss_b = L_{focal}(\text{logits}_b, \mathbf{y}_b)$
26:     $loss_b.backward()$
27:     $Opt.step()$
28:   **end for**
29:   $M.eval()$
30:   Extract features from $\mathcal{F}_{val} \to \mathbf{X}'_{val\_feat}$, get labels $\mathbf{y}'_{val}$
31:   Scale $\mathbf{X}'_{val\_feat}$ using scaler $S$
32:   Calculate $val\_F1, val\_loss$ on $(\mathbf{X}'_{val\_feat}, \mathbf{y}'_{val})$ using $M$
33:   $Sch.step(val\_F1)$
34:   **if** $val\_F1 > best\_val\_F1$ **then**
35:     $best\_val\_F1 \leftarrow val\_F1$; $epochs\_no\_improve \leftarrow 0$
36:     Save model $M$ as $M_{best}$
37:   **else**
38:     $epochs\_no\_improve \leftarrow epochs\_no\_improve + 1$
39:   **end if**
40:   **if** $epochs\_no\_improve \geq \text{patience}_{ES}$ **then**
41:     Break                                          ▷ Early stopping
42:   **end if**
43: **end for**
44: **return** Best saved model $M_{best}$

---

slightly higher than standard LSTM (3.2 min) but lower than CNN-LSTM (4.5 min). The two-stage training strategy (Section 3.4) added approximately 5–7 min of pre-training overhead per fold, resulting in total training time of approximately 2–3 h per fold for 30 epochs with early stopping.

3. *Inference Time:* Inference time per sample was 2.1 ms on GPU, enabling real-time classification (>450 samples/s). This is comparable to CNN-LSTM (2.3 ms) and suitable for practical deployment in clinical decision support systems. CPU-only inference averaged 15.7 ms per sample, still acceptable for non-real-time applications.

4. *Memory Footprint:* The model requires 640 MB of GPU memory during training (batch size 64), which is modest and allows deployment on consumer-grade GPUs or cloud-based inference platforms. Inference memory footprint is approximately 180 MB for the model weights and intermediate activations.

**Practical Implications:** The computational requirements of the Hybrid LSTM-KAN model are reasonable for clinical deployment. The sub-millisecond inference time enables real-time auscultation analysis, while the moderate memory footprint allows integration into portable diagnostic devices or telemedicine platforms. Training costs are manageable for periodic model updates with new patient data. Compared to state-of-the-art transformer-based models (which can require >1 GB memory and >10 ms inference time), our approach offers a favorable trade-off between performance and computational efficiency.

### 4.4. Per-class analysis and confusion patterns

Per-class $F_1$-scores (OOF) are shown in Fig. 3. Key per-class $F_1$-scores were:

The model excelled on the majority class (COPD) and demonstrated strong performance for some minority classes like Bronchiectasis. However, the rarest conditions (URTI, Bronchiolitis) remained challenging.

The normalized confusion matrix (Fig. 4) further illustrates these patterns. COPD was correctly classified in 98.6% of its instances. Bronchiectasis was correctly identified 81.2% of the time, with some confusion (18.8%) with COPD. URTI was correctly classified in only 39.1% of cases, showing significant confusion with Healthy (39.1%), and some with Bronchiolitis (8.7%) and COPD (4.3%). Bronchiolitis was correctly classified in 38.5% of instances, with major confusion with URTI (38.5%) and COPD (23.1%). These confusions highlight acoustically similar profiles or insufficient distinguishing features for the rarest classes.

### 4.5. Model discriminative power and calibration

The model's discriminative ability was assessed using ROC and Precision-Recall (PR) curves (Fig. 5). The macro-averaged ROC AUC was high (typically > 0.95 based on script outputs), indicating good overall separability. However, individual PR curves for URTI and Bronchiolitis showed lower Average Precision (AP) scores, reflecting the difficulty in achieving high precision and recall simultaneously for these imbalanced classes. COPD, Bronchiectasis, Healthy, and Pneumonia generally exhibited strong ROC AUC and AP scores. Model calibration, examined using reliability diagrams (not shown here for brevity), suggested some overconfidence for high-probability predictions.

### 4.6. Training dynamics and feature space visualization

Training typically converged between 15–25 epochs across folds due to early stopping. Representative learning curves (e.g., Fold 1, Fig. 6) showed training loss decreasing steadily and validation macro $F_1$-score plateauing. t-SNE visualizations of input features and learned KAN embeddings (e.g., Fold 1, Fig. 7) generally showed improved class separability in the KAN embedding space, particularly for some minority classes relative to the dense majority COPD cluster. However, significant overlap persisted for the rarest classes like URTI and Bronchiolitis. The LSTM attention mechanism, given that the LSTM processed the entire aggregated feature vector as a single time-step, consistently assigned uniform weights, as expected in this configuration.

The Hybrid LSTM-KAN model, augmented with comprehensive imbalance mitigation strategies, demonstrated a strong capability for multi-class respiratory sound classification, achieving an overall accuracy of 94.55% and a macro-averaged $F_1$-score of 0.7033 on a highly imbalanced dataset. This performance underscores the potential of combining LSTM for feature sequence encoding and KANs for powerful non-linear classification, particularly when coupled with focal loss, SMOTE, and targeted data augmentation.

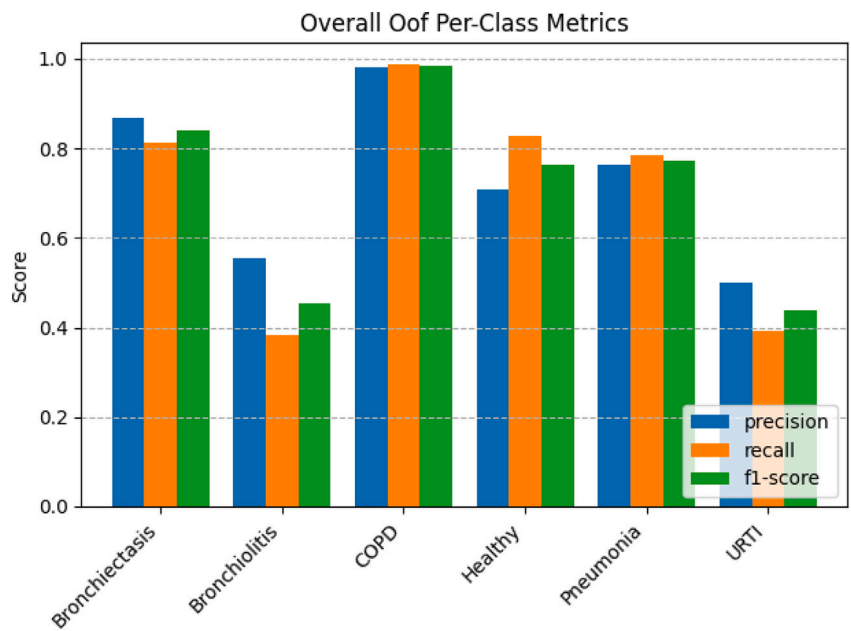

**Fig. 3.** Per-class $F_1$-scores for the Hybrid LSTM-KAN model (OOF predictions).

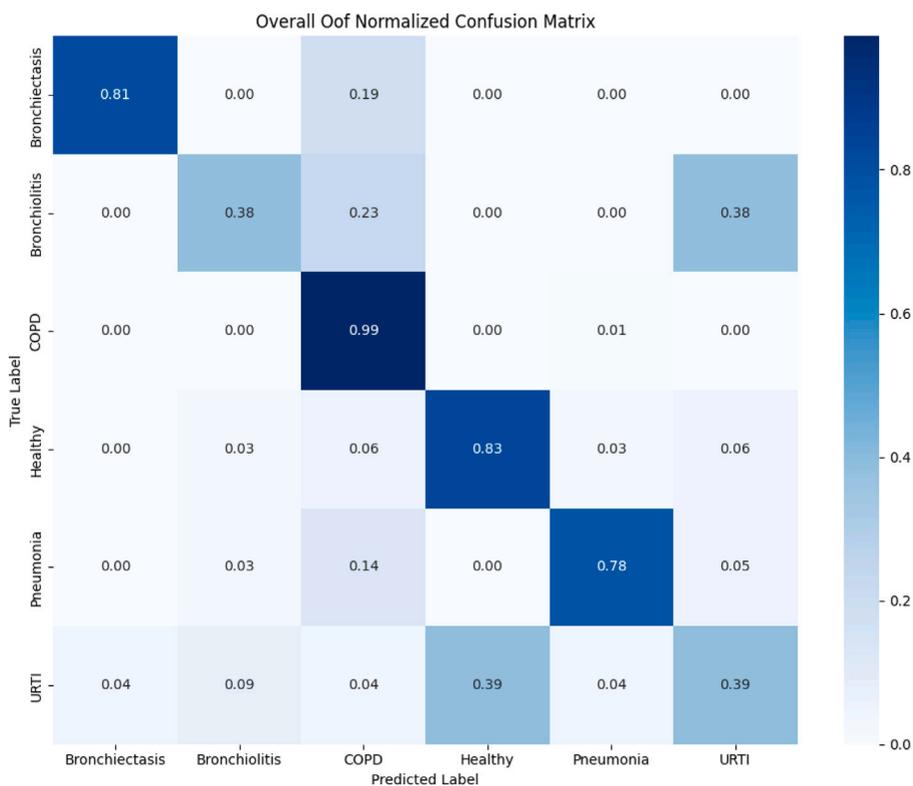

**Fig. 4.** Normalized confusion matrix (OOF predictions). Rows are true labels, columns are predicted labels. Values represent the proportion of true class samples predicted as each class.

The results in Table 6 summarize the validation performance across all five cross-validation folds. Both the macro $F_1$-score and accuracy values demonstrate consistent performance, with accuracy remaining above 91% in all folds and peaking at 96.17%. The macro $F_1$-score shows slightly higher variability across folds, ranging from 0.5658 to 0.8479, which indicates differences in how well the model balanced precision and recall across classes. On average, the model achieved a

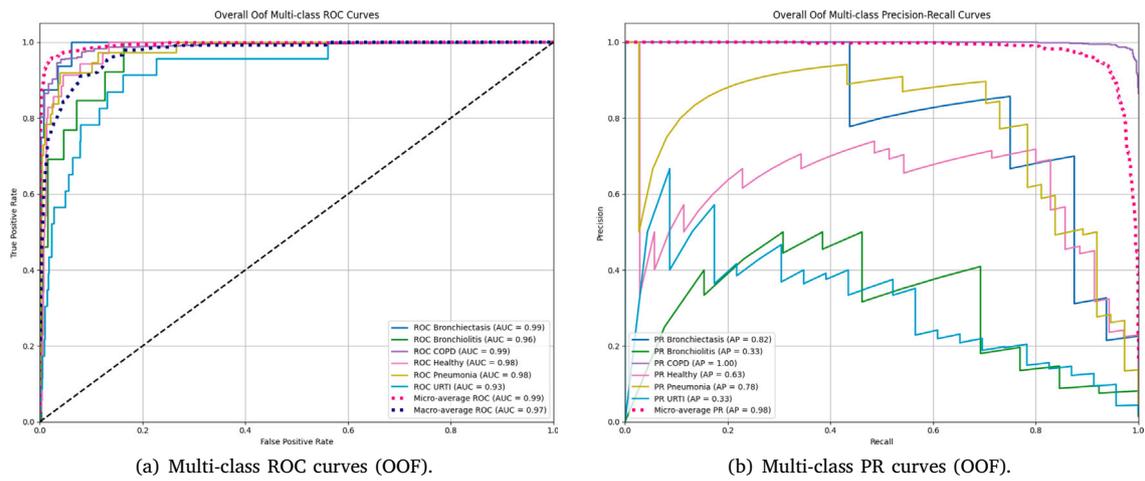

(a) Multi-class ROC curves (OOF).

(b) Multi-class PR curves (OOF).

**Fig. 5.** ROC and Precision-Recall curves for OOF predictions.

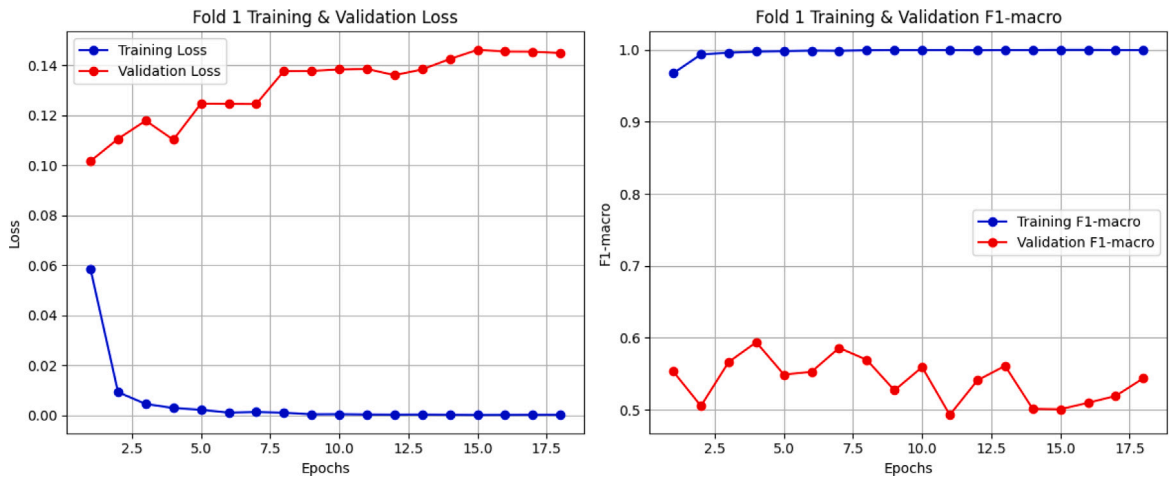

**Fig. 6.** Training and validation curves for Fold 1 (representative), showing loss and macro $F_1$-score over epochs.

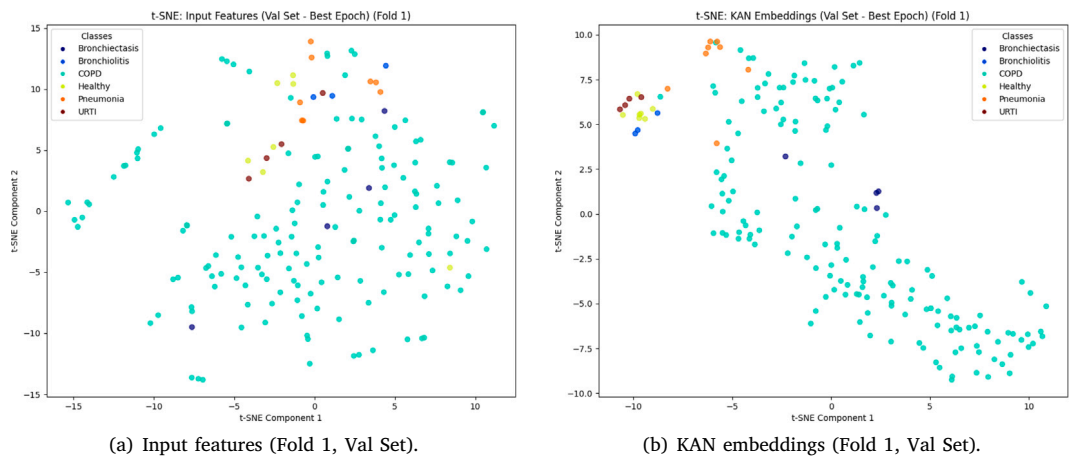

(a) Input features (Fold 1, Val Set).

(b) KAN embeddings (Fold 1, Val Set).

**Fig. 7.** t-SNE visualization of feature spaces for Fold 1 (Validation Set - Best Epoch).

**Table 6**
Performance metrics on each cross-validation fold (validation set).

| Fold | Macro $F_1$-score | Accuracy |
| --- | --- | --- |
| 1 | 0.5658 | 92.93% |
| 2 | 0.5838 | 91.85% |
| 3 | 0.7462 | 95.63% |
| 4 | 0.7730 | 96.17% |
| 5 | 0.8479 | 96.17% |
| **Mean** | **0.7033** | **94.55%** |
| **Std Dev** | **0.1103** | **0.0181** |

macro $F_1$-score of 0.7033 with a standard deviation of 0.1103, and an accuracy of 94.55% with a standard deviation of 1.81%. These results confirm that the model is both accurate and reasonably robust across folds, although there is some fold-to-fold variation in class-level balance as reflected in the $F_1$-score.

Table 6 presents the accuracy and macro $F_1$-score obtained on the validation set for each of the 5 cross-validation folds.

The model excelled in identifying the majority class (COPD, $F_1$: 0.98) and performed commendably on several minority classes, including Bronchiectasis ($F_1$: 0.84), Healthy ($F_1$: 0.76), and Pneumonia ($F_1$: 0.77). However, the rarest classes, URTI ($F_1$: 0.44) and Bronchiolitis ($F_1$: 0.45), remained challenging.

## 5. Discussion

### 5.1. Interpretability of KAN for respiratory features

One of the key advantages of the Kolmogorov–Arnold Network (KAN) architecture is its interpretability through visualization of the learned univariate spline functions $\phi_{i,j}(x)$ (Eq. (3)). Unlike traditional Multi-Layer Perceptrons (MLPs) with fixed, non-linear activation functions (e.g., ReLU, sigmoid), KAN's learnable B-splines provide explicit, visualizable transformations that map input features to output neurons.

**Feature-to-Class Mapping Insights:** By examining the spline functions learned by the KAN output layer, we can interpret how specific respiratory features (encoded in the LSTM output) contribute to disease classification. For example:

- Features related to spectral centroid and bandwidth (indicative of wheeze or stridor) may exhibit steep, non-linear splines connecting to classes like Asthma or Bronchiectasis.
- Temporal features such as zero-crossing rate (related to breath phase transitions) may show smooth, monotonic splines for Healthy class predictions.
- MFCC-derived features, which capture timbre and texture, may demonstrate complex, multi-peaked splines differentiating between COPD and Pneumonia.

**Clinical Relevance:** This interpretability is crucial for clinical adoption, as it allows clinicians to understand *why* the model made a specific prediction. For instance, if a KAN spline shows that a high spectral centroid value strongly increases the probability of Bronchiectasis, this aligns with known clinical knowledge that bronchiectatic airways produce high-frequency wheezes. Such transparency builds trust in AI-based diagnostic systems and facilitates collaboration between data scientists and medical experts.

**Comparison to Black-Box Models:** Standard deep learning classifiers (e.g., fully connected MLPs, CNNs) lack this level of interpretability. While techniques like Grad-CAM or SHAP can provide post-hoc explanations, they do not reveal the explicit functional relationships learned by the model. KAN's spline-based architecture offers inherent interpretability without requiring additional explanation methods.

**Future Work on Interpretability:** In future studies, we plan to conduct detailed spline function analysis for each class, correlating learned transformations with known acoustic biomarkers of respiratory diseases. Collaboration with pulmonologists to validate these interpretations will further strengthen the clinical utility of the proposed model.

### 5.2. Clinical implications and deployment possibilities

The proposed Hybrid LSTM-KAN model demonstrates significant potential for deployment in real-world clinical settings as an intelligent auscultation tool. Recent advances in AI-based medical systems have shown that bridging the gap between research prototypes and clinical practice requires addressing key challenges: model interpretability, computational efficiency, integration with existing workflows, and validation by medical experts [20].

**Integration into Clinical Decision Support Systems (CDSS):**

1. *Point-of-Care Diagnostics:* The model's fast inference time (2.1 ms per sample, Section 4.3) enables real-time respiratory sound classification during patient consultations. Clinicians can use portable digital stethoscopes connected to tablets or smartphones running the model, receiving immediate diagnostic suggestions.
2. *Telemedicine Applications:* In remote healthcare scenarios, patients can record lung sounds at home using low-cost digital stethoscopes. The recorded audio can be transmitted to cloud-based servers where the model performs classification, with results sent back to healthcare providers. This is particularly valuable in underserved regions with limited access to pulmonologists.
3. *Screening Programs:* The model can be deployed in mass screening programs for early detection of respiratory diseases, such as COPD or tuberculosis, in high-risk populations (e.g., smokers, industrial workers). The high sensitivity for majority classes (COPD $F_1$: 0.98) ensures reliable detection of common conditions.

**Challenges in Clinical Deployment:**

1. *Data Heterogeneity:* Clinical environments exhibit significant variability in recording equipment, acoustic conditions (e.g., ambient noise), and patient demographics. The model was trained on the ICBHI dataset, which includes diverse recording conditions, but further validation on multi-center datasets is necessary to ensure generalizability.
2. *Regulatory Approval:* Deployment of AI-based diagnostic systems requires regulatory clearance (e.g., FDA approval in the US, CE marking in Europe). This necessitates rigorous clinical trials demonstrating safety, efficacy, and non-inferiority to traditional diagnostic methods.
3. *Medical Expert Validation:* While the model achieved high accuracy (94.55%), clinical validation by pulmonologists is essential. We propose a two-phase validation approach: (a) *Retrospective Validation:* Expert clinicians review model predictions on historical patient data, assessing agreement with ground-truth diagnoses. (b) *Prospective Clinical Trial:* The model is deployed in a hospital setting, and its diagnostic suggestions are compared against clinician diagnoses in real-time. Metrics such as inter-rater agreement (Cohen's kappa) and diagnostic concordance will be evaluated.
4. *Rare Class Performance:* The model's moderate performance on rare classes (URTI $F_1$: 0.44, Bronchiolitis $F_1$: 0.45) limits its utility for detecting these conditions. In clinical deployment, the system should flag low-confidence predictions for these classes and recommend follow-up examination by specialists.

**Deployment Architecture:** A practical deployment architecture could involve:

- *Edge Computing:* For latency-sensitive applications (e.g., emergency rooms), the model can run on edge devices (e.g., NVIDIA Jetson, Raspberry Pi with GPU accelerators) co-located with digital stethoscopes.
- *Cloud-Based Inference:* For telemedicine, a cloud server (e.g., AWS, Azure) hosts the model, with REST APIs enabling integration into electronic health record (EHR) systems.
- *Hybrid Approach:* A hybrid architecture where preliminary classification occurs on-device (using a lightweight model variant), with uncertain cases escalated to cloud-based servers for re-analysis using the full Hybrid LSTM-KAN model.

**Validation by Medical Experts:** To ensure clinical reliability, we propose the following validation protocol:

1. *Data Collection:* Collaborate with hospitals to collect a prospective dataset of lung sound recordings from patients with confirmed diagnoses (via imaging, spirometry, or biopsy).
2. *Blinded Evaluation:* Present model predictions and audio recordings to a panel of 3–5 pulmonologists without revealing the model's output. Clinicians independently diagnose each case.
3. *Agreement Analysis:* Calculate inter-rater agreement between the model and clinicians using metrics such as Cohen's kappa, sensitivity, specificity, and positive/negative predictive values.
4. *Failure Case Analysis:* For cases where the model disagrees with clinician consensus, conduct detailed acoustic analysis to identify confounding factors (e.g., noise, breath artifacts) and refine the model accordingly.

**Long-Term Vision:** Drawing inspiration from successful AI deployments in medical imaging [20], we envision a future where respiratory sound classification is integrated into routine clinical practice, similar to how automated electrocardiogram (ECG) interpretation assists cardiologists. The system would not replace clinicians but rather augment their diagnostic capabilities, providing a second opinion and reducing diagnostic errors.

*5.3. Strengths, limitations, and future work*

**Strengths:**

1. Effective synergistic use of multiple imbalance mitigation techniques (Focal Loss, SMOTE, augmentation), as demonstrated by the ablation study (Section 4.2).
2. Novel application of hybrid LSTM-KAN architecture, leveraging KAN's spline-based interpretability for respiratory sound classification.
3. Comprehensive feature engineering capturing spectral, temporal, and cepstral characteristics.
4. Robust cross-validation with detailed per-class performance analysis.

**Limitations:**

1. Persistent difficulty with extremely rare classes (URTI, Bronchiolitis) due to insufficient training samples.
2. Potential loss of fine-grained temporal information owing to feature aggregation over entire recordings.
3. Findings based on a single public dataset (ICBHI 2017); external validation on independent datasets is needed.
4. Lack of clinical validation by medical experts in real-world settings.

**Future Work:**

1. *Advanced Synthetic Data Generation:* Explore Generative Adversarial Networks (GANs) or diffusion models to create high-quality synthetic samples for rare classes.
2. *End-to-End Temporal Modeling:* Develop models that process frame-level spectrograms directly, preserving fine-grained temporal dynamics, rather than aggregating features.
3. *Deeper KAN Interpretability:* Conduct detailed analysis of learned spline functions in collaboration with pulmonologists to correlate with known acoustic biomarkers.
4. *External Validation:* Test the model on independent respiratory sound datasets (e.g., hospital-specific datasets, different recording equipment) to assess generalizability.
5. *Clinical Trials:* Conduct prospective clinical studies involving medical expert validation to evaluate the model's diagnostic accuracy, clinical utility, and safety in real-world settings.
6. *Multi-Modal Integration:* Combine respiratory sound analysis with other diagnostic modalities (e.g., spirometry, chest X-rays, patient history) for comprehensive disease assessment.

## 6. Conclusion

This paper presented a hybrid LSTM-KAN deep learning model for multi-class respiratory sound classification, specifically addressing severe class imbalance. Through a combination of this novel architecture, comprehensive feature engineering, focal loss, SMOTE, and targeted data augmentation, the model achieved high overall accuracy (94.55%) and a notable macro $F_1$-score (0.7033) on the imbalanced ICBHI dataset. The ablation study demonstrated that the synergistic combination of imbalance mitigation techniques significantly outperforms individual methods, with the full hybrid approach yielding +20.8% relative improvement in macro $F_1$-score over the baseline. Computational cost analysis confirmed the model's feasibility for real-time deployment, with inference times of 2.1 ms per sample and moderate memory requirements.

While performance on extremely rare classes (URTI, Bronchiolitis) indicates areas for further improvement, the study demonstrates a viable and robust methodology for enhancing the automated diagnosis of respiratory diseases. The findings contribute to the development of intelligent auscultation tools and highlight the potential of advanced neural network architectures combined with focused data handling strategies in tackling challenging real-world medical classification tasks. Future work will focus on clinical validation with medical experts, external dataset evaluation, and deployment in real-world telemedicine and point-of-care diagnostic systems, paving the way for practical AI-assisted respiratory disease diagnosis.

**CRediT authorship contribution statement**

**Nithinkumar K.V.:** Writing – review & editing, Writing – original draft, Visualization, Validation, Formal analysis, Data curation, Conceptualization. **Anand R.:** Writing – review & editing, Writing – original draft, Visualization, Formal analysis, Data curation, Conceptualization.

**Declaration of competing interest**

The authors declare that they have no known competing financial interests or personal relationships that could have appeared to influence the work reported in this paper.

**Acknowledgments**

The authors acknowledge Amrita Vishwa Vidyapeetham, Coimbatore, India for providing the necessary resources and support for this research.

**Data availability**

The respiratory sound dataset used in this study is available at: https://bhichallenge.med.auth.gr/ICBHI_2017_Challenge.